\documentclass{Interspeech}

\interspeechcameraready

\title{Adapting Whisper for Lightweight and Efficient Automatic Speech Recognition of Children for On-device Edge Applications}

\author[affiliation={1}]{Satwik}{Dutta}
\author[affiliation={1}]{Shruthigna}{Chandupatla}
\author[affiliation={1}]{John}{H.L. Hansen}

\affiliation{Center for Robust Speech Systems}{The University of Texas at Dallas}{USA}
\email{satwik.dutta@utdallas.edu, shruthigna.chandupatla@utdallas.edu, john.hansen@utdallas.edu}

\usepackage{comment,subcaption}

\begin{document}

\maketitle

\begin{abstract}
    
    Reliability on cloud providers for ASR inference to support child-centered voice-based applications is becoming challenging due to regulatory and privacy challenges. Motivated by a privacy-preserving design, this study aims to develop a lightweight \& efficient Whisper ASR system capable of running on a Raspberry Pi. Upon evaluation of the MyST corpus and by examining various filtering strategies to fine-tune the `tiny.en' model, a Word Error Rate (WER) of 15.9\% was achieved (11.8\% filtered). A low-rank compression reduces the encoder size by 0.51M with 1.26$\times$ faster inference in GPU, with 11\% relative WER increase. During inference on Pi, the compressed version required $\approx$2 GFLOPS fewer computations. The RTF for both the models ranged between [0.23-0.41] for various input audio durations. Analyzing the RAM usage and CPU temperature showed that the PI was capable of handling both the tiny models, however it was noticed that small models initiated thermal throttling.   
\end{abstract}

\section{Introduction}
\label{sec:intro}

The focus on developing Automatic Speech Recognition (ASR) systems for children has accelerated over the years, given the enormous benefit of its application in educational technology \cite{hembise21_interspeech,okur-etal-2022-end,kelly20b_interspeech}, spoken language learning \cite{piton23_interspeech}, games/toys/robots \cite{kennedy2017child}, and speech/language development \cite{dutta22_interspeech,dutta-etal-2022-activity} of children. This technology is at the core of various `AI-based products' for children, at home or in the classroom. Most of these products often rely on third-party cloud service providers for voice-driven applications, which include uploading the voice file to the cloud server, performing transcription, and then being followed by the intended use/application. It is expected that the voice file would be deleted after use or after a given time, as per the law (of the country/state). In the U.S., such a service provider (or company) is bound by the Children’s Online Privacy Protection Act (COPPA) enforced by the Federal Trade Commission (FTC). As per a comprehensive report published by the FTC \cite{federal2024ftc} and numerous lawsuits \cite{federalcase},
in the recent years several companies offering products in gaming, social media, and educational technology have violated the COPPA enforcement policy related to the collection and use of personal information of children (including voice \cite{federalalexa}).
Therefore, even with their enormous benefits, child-centered technology products come with their challenges, such as the risk of violating children's privacy \cite{dutta24_syndata4genai,DUTTA2025103460}.  

A solution to this challenge would be to rely on a local or on-device ASR backend \cite{oh2021device,10.1145/3643832.3661886}, which works offline, where the voice data stays on the device and is deleted as soon as the transcripts are generated, and is cost-effective. This ensures a privacy-preserving design without compromising COPPA (US), or equivalent privacy laws such as GDPR (Europe) \cite{gdpr} and DPA (UK)\cite{dpa}. Such a design also builds confidence among parents, teachers, and practitioners to use the product. With this motivation, this study aims to develop a lightweight and efficient ASR system for children that is capable of running locally on an edge device such as a Raspberry Pi.  
Children's developing spoken skills, knowledge of grammar, and changing vocal characteristics \cite{gerosa2007acoustic} add challenges to developing a robust child ASR system. This is still considered a low-resource scenario when compared with that of adults, with the `My Science Tutor Children's Conversational Speech Corpus' (MyST) \cite{ward2019my} being the largest American English-based child ASR corpus available to the research community to date. Prior studies \cite{shivakumar2022end,jain23_interspeech,kid-whisper,fan24b_interspeech} using the MyST corpus relied on various state-of-the-art ASR systems, including Whisper \cite{pmlr-v202-radford23a}. However, to the best of our knowledge, most prior studies were focused on achieving robust ASR systems (lowest word error rate) rather than the ASR system's computational or memory efficiency. Although, a recent study \cite{DUTTA2025103460} using the MyST corpus claimed that using discrete speech units for ASR achieves similar performance to a much larger (x10) state-of-the-art End-to-End ASR system using WavLM \cite{chen2022wavlm} features. 

In this study, we primarily focus on adapting Whisper ASR models using the MyST corpus for evaluation. First, we explore the best lightweight ASR model using various data filtering strategies, and further compress it to reduce inference time with a slight compromise in transcription performance. Second, we perform model inference analysis on-device on a Raspberry Pi where we analyze: audio recording quality, computational efficiency in GFLOPS and RTF, overhead analysis with RAM usage and CPU temperature. Model checkpoints,  recipes, and the full list of the test split utterances are available on GitHub\footnote{https://github.com/SatwikDutta/LiteChildASR}.

\begin{table*}[th]
  \caption{Summary of the MyST splits for three filtering strategies in terms of the amount of audio and WER for Whisper Large V2.}
  \label{tab:myst-summary}
  \centering
  \begin{tabular}{l|ccc|c|cc|cc}
    \toprule
    \multicolumn{1}{c|}{\textbf{Data version}} & \multicolumn{3}{c|}{\textbf{Filtering}} & \textbf{Train} & \multicolumn{2}{c|}{\textbf{Dev}} & \multicolumn{2}{c}{\textbf{Test}} \\
    \cline{2-4} \cline{6-7} \cline{8-9}
    & \textbf{F1} & \textbf{F2} & \textbf{F3} & \textbf{hrs [\#]} & \textbf{hrs [\#]} & \textbf{WER} & \textbf{hrs [\#]} & \textbf{WER} \\
    \midrule
    \midrule
    A - Original & $\times$ & $\times$ & $\times$ & 143.7 & 22.8 & 16.4 & 28.0 & 19.7\\
    B - Filtered Original & \checkmark & \checkmark & $\times$ & 125.3 & 19.9 & 11.6 & 23.7 & 13.0\\
    C - 30s version & $\times$ & $\times$ & \checkmark & 223.6 & 36.2 & 15.4 & - & - \\
    D - Filtered 30s vers. & \checkmark & \checkmark & \checkmark & 194.9 & 21.3 & 11.5  & - & - \\
    \bottomrule
    \multicolumn{9}{l}{$F1\rightarrow$ All utterances with more than 50\% WER using OpenAI Whisper Large V2 are discarded.} \\
    \multicolumn{9}{l}{ 
    $F2\rightarrow$ All utterances with less than 3 words are discarded.} \\
    \multicolumn{9}{l}{$F3\rightarrow$ Utterances within a given session are combined to create 25-30 second samples.}\\
  \end{tabular}
  \vspace{-15pt}
\end{table*}
\vspace{-5pt}
\section{MyST: The My Science Tutor Children’s Conversational Speech Corpus}
\label{sec:corpus:myst}

The `My Science Tutor Children's Conversational Speech Corpus' or MyST corpus \cite{ward2019my},
contains approximately 210 hrs of transcribed children's utterances (470 hrs in total) during conversation between 1371 students in grades 3-5 (ages 8-11) and a virtual science tutor on various topics such as Physics, Chemistry, Mathematics, Geography, Biology, Geology, etc. 
This study used the same training, development, and testing splits as the original corpus. The majority of the utterances in the MyST corpus were less than 10 seconds. By manually analyzing random utterances across the corpus, we noticed various transcription errors (incorrect non-linguistic markers, incorrect sentences, etc) and audio quality issues (speaking close to the microphone, speaking softly, etc). 

\noindent \textbf{Filtering Strategies}: 
 A summary of the MyST data versions by applying various filtering strategies on the training, development, and test splits is highlighted in Table \ref{tab:myst-summary}. All utterances over 30 seconds were discarded from the training and development splits, since we primarily use Whisper ASR \cite{pmlr-v202-radford23a} with a 30-second receptive field. Utterances
 across all splits containing no signal, only silence or background noise were discarded across all splits. All the ground-truth transcriptions were normalized using WhisperTokenizer \cite{wt}
 . This resulted in the data version `\textbf{A - Original}'. By discarding all utterances with more than 50\% WER using OpenAI Whisper Large V2 \cite{hfo} (F1)
 and with less than 3 words in the ground truth transcriptions (F2), the data version `\textbf{B - Filtered Original}' was obtained.  The 50\% WER threshold was used to discard utterances with incorrect transcriptions. Utterances with fewer than 3 words were discarded, as these did not have the necessary context especially to differentiate between homophones (`right' vs. `write'). Due to the 30-second receptive field and the original distribution of the utterances in training and development (with the majority of short utterances), a third filtering strategy was adopted to avoid padding shorter utterances during training. In this filtering step (F3), all utterances within a given ‘data collection session’
 , in the training and development splits (of data version `\textbf{A - Original}'), were only used if their duration was between 25-30 seconds or concatenated to create 25-second to 30-second utterances. This resulted in the data version `\textbf{C - 30s version}'. Finally, as data version `A - Original' was filtered to create 'B - Filtered Original', similarly 'C - 30s version' was filtered to create `\textbf{D - Filtered 30s vers.}'. These three filtering steps were primarily done to match (to the best of our ability) the training/evaluation data with prior work \cite{kid-whisper} and compare our results. In total (as shown in Tab.\ref{tab:myst-summary}), 4 versions of the training and development splits and 2 versions of the testing split were created for this study. 
\vspace{-7pt}
\section{Experimental Details}
\label{sec:exp}

\subsection{Model Training and Evaluation (on GPU)}
\label{sec:exp:mdltrn}

All training (fine-tuning) and evaluation recipes were adapted from HuggingFace and OpenAI.
For zero-shot evaluation, all models were downloaded from OpenAI's Hugging Face model checkpoints \cite{hfo}.
These include both English and multilingual versions of \textit{tiny}, \textit{base}, \textit{small}, and \textit{medium}, and version 3 of \textit{large} and \textit{large turbo}. For model fine-tuning, all 4 data versions (Tab.\ref{tab:myst-summary}) were considered. The batch size was set to 128 for \textit{tiny}, \textit{base}, and \textit{small}, and to 64 and 16 for \textit{medium} and \textit{large} models, respectively. Fine-tuning was performed until convergence (with a patience value of 5), and the best checkpoints based on the WER metric were used for evaluation. All experiments for Sec.\ref{sec:exp:mdltrn} were performed on an Nvidia A100 40GB GPU.

Distil-Whisper \cite{gandhi2023distil}, leveraged pseudo labelling to distill the Whisper model into a smaller variant by reducing the number of decoder
layers from 32 to 2 for Large-v3. Additional research \cite{distillhf} on Distil-Whisper suggests that it might be quite difficult to retain performance for smaller models, and additionally \textit{distil-small.en} slower than \textit{distil-large-v2}. Given this challenge, we did not consider Distil-Whisper for our study. Instead a low-rank compression scheme \cite{kamahori2025liteasr} was adopted. This study proposed a low-rank compression scheme for ASR encoders that significantly reduces inference costs while maintaining transcription accuracy. This study achieved similar transcription accuracy by compressing Whisper large-v3’s encoder size by over 50\%, matching  Whisper medium’s size. 
\begin{figure}[h]
  \centering
  \includegraphics[width=\linewidth]{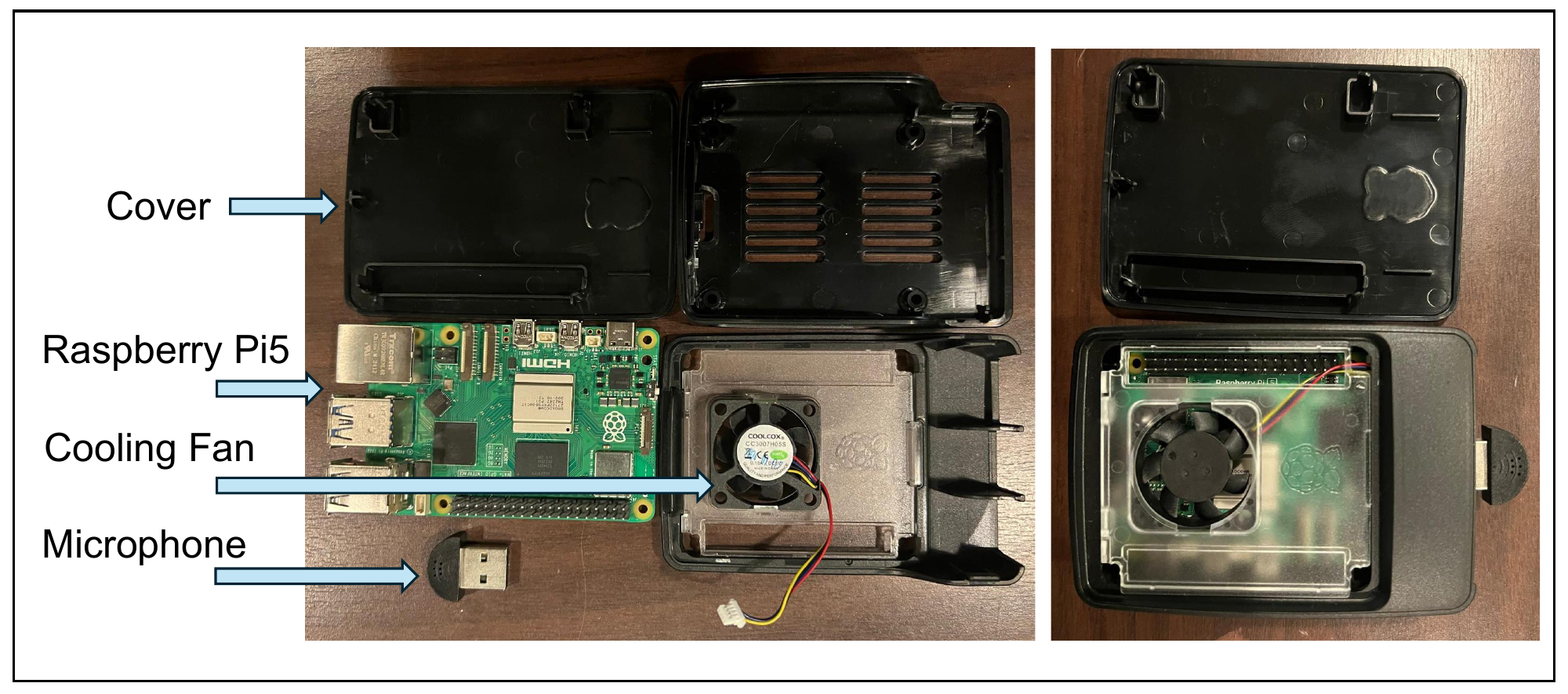}
  \vspace{-15pt}
  \caption{Raspberry PI and accessories.}
  \label{fig:pi}
  \vspace{-15pt}
\end{figure}

\begin{figure}[h]
  \centering
  \includegraphics[width=\linewidth]{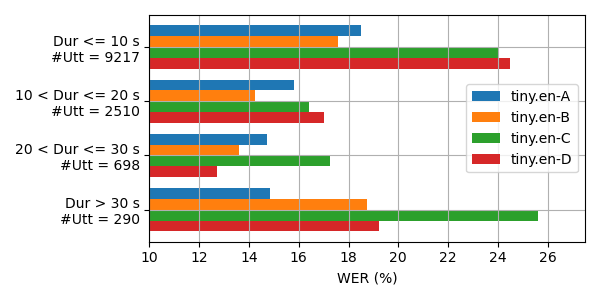}
  \vspace{-25pt}
  \caption{Comparison of A(Org) Test WER by audio duration.}
  \label{fig:werbysize}
  \vspace{-20pt}
\end{figure}
\begin{figure}[h]
  \centering
  \includegraphics[width=\linewidth]{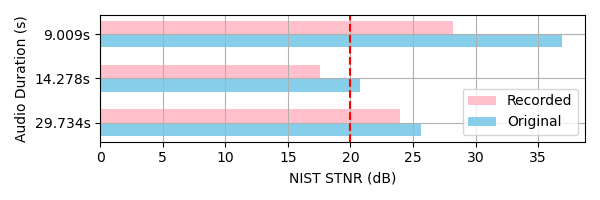}
  \vspace{-25pt}
  \caption{Comparison of NIST STNR: Original vs. Recorded.}
  \label{fig:org_vs_rec:niststnr}
  \vspace{-20pt}
\end{figure}
\vspace{-10pt}
\begin{figure}[h]
  \centering
  \includegraphics[width=\linewidth]{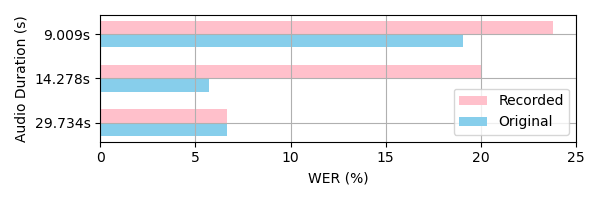}
  \vspace{-25pt}
  \caption{Comparison of WER (\%) for the \textit{tiny.en-`A'} model: Original vs. Recorded.}
  \label{fig:org_vs_rec:wer}
  \vspace{-20pt}
\end{figure}
\begin{figure*}[h]
  \centering
  \includegraphics[width=\linewidth]{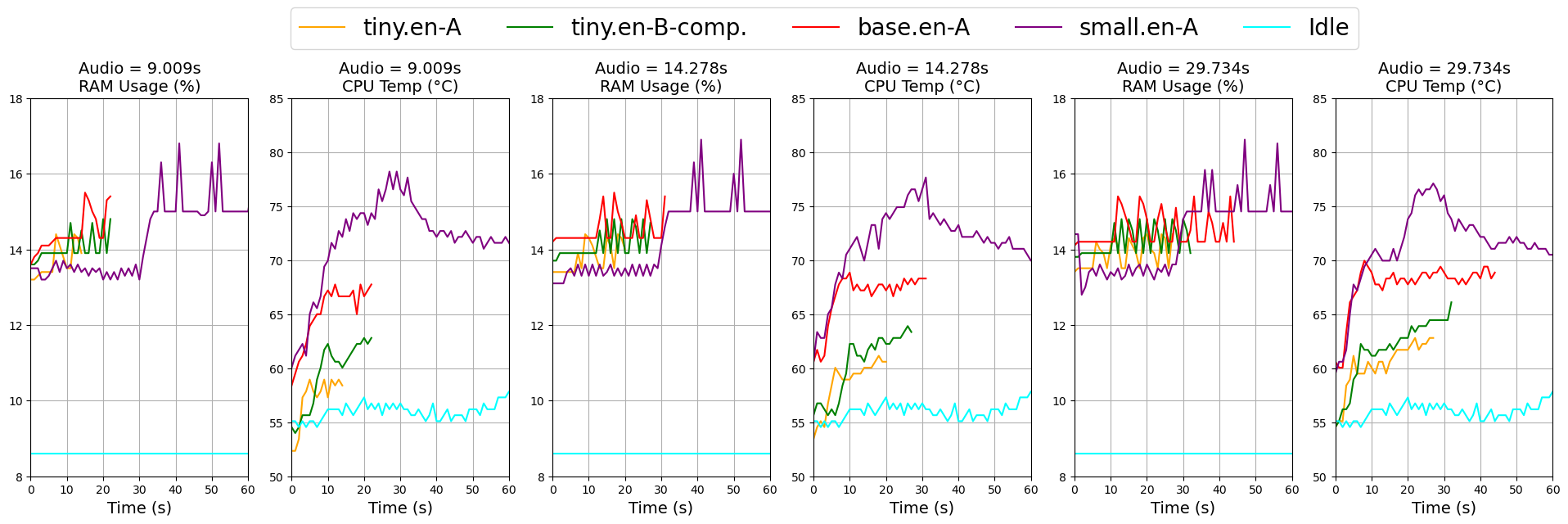}
  \caption{RAM Usage Percentage and CPU Temperature (this includes pre-processing steps like loading of model, audio, etc.).}
  \label{fig:ramcpu}
  \vspace{-15pt}
\end{figure*}
\vspace{10pt}
\subsection{Model Inference on the Raspberry Pi}
\label{sec:exp:mdlinf}
Raspberry Pi 5 and other accessories are shown in Fig.\ref{fig:pi}. 3 random test utterances were chosen from the MyST corpus and also recorded in a sound booth using a speaker (at 1.57m) to imitate a `speaking child'. For each recorded input audio, the NIST STNR \cite{niststnr}
and the WER based on the best lightweight model output were compared with that of the original input audio. The recorded audio was downsampled to 16 kHz, similar to the original input audio. For each model and input audio pair, we calculate the \textit{Transcription Processing Time} or \textit{Latency}. For our case, this also includes feature extraction and text decoding from predicted tokens. However, this does not include any pre-processing time for loading the model or audio load time. We primarily measure: (1) \textit{Giga Floating Point Operations (GFLOPS)}: to measure the model computational complexity,  and (2)\textit{ Real-time factor (RTF)}: to measure the overall ASR system efficiency. An RTF less than 1.0 indicates the system operates faster than real-time. Additionally, we measure: (3) \textit{RAM Usage Percentage}: to measure the system memory usage during transcription, and (4) \textit{CPU temperature} ($^{\circ}$C): to check whether any processing on the PI reaches between the thermal throttling temperature of 80°C and the critical temperature of 85°C. These analyses were repeated ten times. 
\vspace{-5pt}

\subsection{Model Training and Evaluation (on GPU)}
\label{sec:results:mdltrn}

\subsubsection{Zero-shot vs. Fine-tuned}
\label{sec:results:mdltrn:finetune}

All results for this section are shown in Table \ref{tab:myst-train-result}. Irrespective of the test data version, the zero-shot test WER reduced as the model size increased from \textit{tiny} to  \textit{medium}, and saturated thereafter. 
Fine-tuning reduced the relative \textbf{A(Org)} test WER by 36\% for \textit{small.en}, 49\% for \textit{base.en}, and 55\% for \textit{tiny.en} respectively. 
Prior work achieved the best test WER of 9.11\% \cite{kid-whisper} (25.8 hrs) and 8.8\% \cite{fan24b_interspeech} (25 hrs, $>30$s discarded) for similar filtering as our \textbf{B (Fil)} using \textit{small.en}, close to our WER of 8.9\%. 
For \textit{tiny.en} particularly, the performance depended on the type of training data. Our results indicated that using only 25-30s utterances (filtering strategy F3) did not aid performance. A further analysis by utterance length (Fig.\ref{fig:werbysize}) showed that such models (C,D) did not perform well for shorter utterances, and the filtering strategy F1 and F2 only worked for longer utterances. On the contrary, the filtering strategy F1 and F2 aided performance for models trained on the original data (A $\rightarrow$ B) for utterances less than or equal to 30-second duration. Overall, the smallest model to achieve the best \textbf{A(Org)} test WER of 15.9\% was \textbf{tiny.en-`A'}. Our experiments also showed that English models performed much better than multilingual ones, and fine-tuning on \textit{large} models degraded performance (insufficient data). We consider the test data version \textbf{A(Org)} to be more robust and challenging than \textbf{B (Fil)}, therefore, all further results use data version \textbf{A(Org)}. 

\begin{table}[th]
  \caption{Model Evaluation: Zero-shot vs. Fine-tuned.}
  \label{tab:myst-train-result}
  \centering
  \begin{tabular}{l|c|l|c|c}
    \toprule
    \multicolumn{3}{c|}{\textbf{Model}} & \multicolumn{2}{c}{\textbf{Test WER (\%)}} \\
    \midrule 
    \multicolumn{1}{c|}{\textbf{Size(.lang)}} & \multicolumn{1}{c|}{\textbf{\#Parm}} & \textbf{Finetuned on} & \textbf{A (Org)} & \textbf{B (Fil)} \\
    \midrule
    & & $\times$ Zero-shot & 28.0 & 18.2 \\
    & & \checkmark A (Org) & \textbf{15.9} & 11.8 \\
    tiny.en & 39M & \checkmark B (Fil) & 16.6 & \textbf{11.2} \\
    & & \checkmark C (Org-30s) & 20.4 & 12.2 \\
    & & \checkmark D (Fil-30s) & 19.5 & 12.4 \\
    \midrule
    tiny & 39M & $\times$ Zero-shot & 31.4 & 23.1 \\
    \midrule
    \midrule
    & & $\times$ Zero-shot & 22.9 & 15.7 \\
    & & \checkmark A (Org) & \textbf{13.9} & \textbf{9.9} \\
    base.en & 74M & \checkmark B (Fil) & 15.6 & 10.0 \\
    & & \checkmark C (Org-30s) & 16.1 & 10.7 \\
    & & \checkmark D (Fil-30s) & 17.8 & 10.9 \\
    \midrule
    base & 74MM & $\times$ Zero-shot & 25.6 & 17.4 \\
    \midrule    
    \midrule
    & & $\times$ Zero-shot & 18.8 & 12.9 \\
    & & \checkmark A (Org) & \textbf{13.0} & 9.9 \\
    small.en & 244M & \checkmark B (Fil) & 13.7 & \textbf{8.9} \\
    & & \checkmark C (Org-30s) & 16.0 & 11.5 \\
    & & \checkmark D (Fil-30s) & 14.4 & 9.9 \\
    \midrule
    small & 244M & $\times$ Zero-shot & 20.4 & 14.3 \\
    \midrule
    \midrule
    medium.en & 769M & $\times$ Zero-shot & 18.2 & 12.9 \\
    medium & 769M & $\times$ Zero-shot & 17.9 & 12.4 \\
    large-v3 &  1550M & $\times$ Zero-shot & 17.6 & 12.7 \\
    large-v3-tb & 809M & $\times$ Zero-shot & 18.9 & 13.0 \\
    \bottomrule
  \end{tabular}
  \vspace{-15pt}
\end{table}

\vspace{-10pt}
\section{Results and Discussion}
\label{sec:results}
\subsubsection{Evaluation of Low-Rank Compression}
\label{sec:results:mdltrn:lowrank}

All results for this section are shown in Tab.\ref{tab:myst-low-rank}. By comparing OpenAI's multilingual \textit{tiny} model with the compressed version \cite{kamahori2025liteasr}, the relative test WER increases by 8.24\% for a reduction of 0.22M encoder parameters and quicker inference time on GPU. Particularly for fine-tuned \textit{tiny.en}, two compressed models were chosen. Both of these compressed models were calibrated with 500 samples and data compression threshold of $0.999$ for both the self-attention and MLP layers. By compressing \textit{tiny.en-`A'}, the relative test WER increases by 19.32\% relative for a reduction of 0.55M encoder parameters and a reduction in inference time by a factor of 0.79. A similar reduction in inference time was observed by compressing \textit{tiny.en-`B'}, which increased the relative test WER by 11.36\% for a reduction of 0.51M encoder parameters. 

\vspace{-1.0mm}
\begin{table}[th]
  \caption{Evaluation of Low-Rank Compression (LRC).}
  \label{tab:myst-low-rank}
  \centering
  \begin{tabular}{l|l|c|c|c|c}
    \toprule
    \multicolumn{4}{c|}{\textbf{Model}} & \multicolumn{2}{c}{\textbf{A(Org) Test}} \\
    \midrule 
    \multicolumn{1}{c|}{\textbf{Size}} & \multicolumn{1}{c|}{\textbf{Fine-}} & \textbf{LRC} & \textbf{\#Parm} & \textbf{WER} & \textbf{Normali} \\
    \multicolumn{1}{c|}{\textbf{(.lang)}} & \multicolumn{1}{c|}{\textbf{tune}} &  & \textbf{Enc(M)} & \textbf{(\%)} & \textbf{zed Time} \\
    \midrule
    tiny &  $\times$ & $\times$ & 7.63 & 31.4 & 1.06 \\
    tiny \cite{kamahori2025liteasr} & $\times$ & \checkmark & 7.41 & 34.1 & 0.83 \\
    \midrule
     & $\times$ & $\times$ & 7.63 & 28.0 & 1.01 \\
     & \checkmark A & $\times$ & 7.63 & \textbf{15.9} & \textbf{1.00} \\
    tiny.en & \checkmark A & \checkmark & 7.08 & \fbox{19.3} & \fbox{0.79} \\
      & \checkmark B & $\times$ & 7.63 & 16.6 & - \\
      & \checkmark B & \checkmark & 7.12 & \fbox{18.6} & \fbox{0.79} \\
    \midrule
    \midrule
    base.en  & \checkmark A & $\times$ & 19.82 & \textbf{13.9} & 1.10 \\
    small.en & \checkmark A & $\times$ & 87.00 & \textbf{13.0} & 1.36 \\
    \bottomrule
    \multicolumn{6}{l}{\textbf{\#Param Dec.(M)}: tiny$\rightarrow$29.55, base$\rightarrow$52.00, small$\rightarrow$153.58}
  \end{tabular}
  \vspace{-3.0mm}
\end{table}
\vspace{-3.0mm}
\begin{table}[th]
  \caption{Computational Efficiency: GFLOPS and Average RTF over 10 repeated runs. (RTF variance values shown in GitHub)}
  \label{tab:computational efficieny}
  \centering
  \begin{tabular}{l|l|c|c|c|c}
    \toprule
    \textbf{Audio} & \multicolumn{3}{c|}{\textbf{Model}} & \textbf{GFLOPS} & \textbf{RTF} \\
    \cline{2-4}
    \textbf{Dur(s)} & \multicolumn{1}{c|}{\textbf{Size}} & \multicolumn{1}{c|}{\textbf{Fine-}} & \textbf{LRC} &  &  \\
     &\multicolumn{1}{c|}{\textbf{(.lang)}} & \multicolumn{1}{c|}{\textbf{tune}} &  &  &  \\
    \midrule
    &  & \checkmark A & $\times$ & 26.24 & 0.39 \\
    & tiny.en & \checkmark A & \checkmark & \textbf{24.25} & \textbf{0.31} \\
    9.00 &  & \checkmark B & \checkmark & 24.71 & 0.41 \\
    \cline{2-6}
    & base.en  & \checkmark A & $\times$ & 68.01 & 0.61 \\
    & small.en & \checkmark A & $\times$ & 303.69 & 1.88 \\
    \hline
    \hline
    &  & \checkmark A & $\times$ & 27.03 & 0.36 \\
    & tiny.en & \checkmark A & \checkmark & \textbf{25.15} & \textbf{0.33} \\
    14.28 &  & \checkmark B & \checkmark & 25.56 & 0.37 \\
    \cline{2-6}
    & base.en  & \checkmark A & $\times$ & 69.76 & 0.61 \\
    & small.en & \checkmark A & $\times$ & 307.86 & 1.66 \\
    \hline
    \hline
    &  & \checkmark A & $\times$ & 28.16 & \textbf{0.23} \\
    & tiny.en & \checkmark A & \checkmark & \textbf{26.50} & 0.25 \\
    29.73 &  & \checkmark B & \checkmark & 26.63 & 0.25 \\
    \cline{2-6}
    & base.en  & \checkmark A & $\times$ & 71.71 & 0.41 \\
    & small.en & \checkmark A & $\times$ & 312.86 & 1.05 \\
    \bottomrule
  \end{tabular}
\end{table}
\subsection{Model Inference on the Raspberry Pi}
\label{sec:results:mdlinf}

\subsubsection{Original vs. Recorded Audio}
\label{sec:results:mdlinf:audqua}

The room acoustics was measured using Decibel X Pro Sound Meter app
, with the sound pressure level ranging between 28 and 30 dBA in quiet conditions, and between 80 and 90 dBA when the input audios were being played using the speaker. The comparison of NIST STNR and WER of the original and recorded audios is shown in Fig.\ref{fig:org_vs_rec:niststnr} and Fig.\ref{fig:org_vs_rec:wer} respectively. The NIST STNR value of the recorded audios was lower than that of the equivalent original audios. Except for the \textit{14.28s} sample, all the audios had a NIST STNR value greater than 20 dB. This degradation in the audio quality also impacts the WER (higher) for this sample. One of the reasons for this degradation in audio quality may be due to the microphone used for recording. Although the recorded audio did not miss any spoken words, slight system noise (such as that of the fan) was captured. For further analyses, we only use the original input audios.  
\vspace{-8pt}
\subsubsection{Computational Efficiency: GFLOPS and RTF}
\label{sec:results:mdlinf:compeff}

All results for this section are shown in Tab.\ref{tab:computational efficieny}. It was observed that the GFLOPS gradually increased with an increase in audio length for a given model. Both the compressed versions of fine-tuned \textit{tiny.en} required fewer computations ($\approx$2 GFLOPS) than un-compressed fine-tuned \textit{tiny.en}. With increasing audio length and same model size, RTF reduced in most cases. While the fine-tuned \textit{tiny.en} had an RTF between 0.23 and 0.39, the RTF ranged between 0.25 to 0.41 for the compressed ones. 
\vspace{-8pt}
\subsubsection{RAM usage and CPU temperature}
\label{sec:results:mdlinf:ramcpu}

All results for this section are shown in Figure \ref{fig:ramcpu}. During the processing of each input audio by the different models, the RAM usage and CPU temperature were tracked in parallel. Apart from the transcription, the processing included operations such as loading the respective model and audio, and saving timestamps. Lighter models finish quicker, as evident from the figure. The overall RAM usage for the fine-tuned \textit{tiny.en-`A'}, including the compressed one \textit{tiny.en-`B'-compressed}, stayed within 15\%. For the fine-tuned \textit{base.en-`A'} and \textit{small.en-`A'} models, the RAM usage increased to 15.5\% and 17\% maximum, respectively. For processing of the \textit{tiny.en-`A'}, \textit{tiny.en-`B'-compressed}, and \textit{base.en-`A'}, the CPU temperature remained below 80 to 85°C, i.e., the thermal throttling to the critical range. However, for the \textit{small.en-`A'} model, it was observed that when the temperature reaches closer to 80°C, thermal throttling of the Pi was activated (temperature reduces gradually, potentially by switching on the fan). It is known that the thermal throttling impacts the processing (speed) of the Raspberry Pi, adding additional overhead. Therefore, the \textit{small.en-`A'} model might not be suitable for a Raspberry Pi of a similar configuration.
\vspace{-10pt}
\section{Conclusion}
In this study, we developed a lightweight and efficient Whisper ASR model for running on-device on a Raspberry Pi. 
Our lightweight fine-tuned `tiny.en' model achieves a 15.9\% WER for the MyST corpus.
It also demonstrates our analysis on various filtering strategies, with a conclusion that using only 25-30s utterances for training/development degrades model performance, building on prior work \cite{kid-whisper}. 
Our efficient model, generated by low-rank compression, resulted in lower computation and faster inference.
Inference analysis of the lightweight and efficient models showed that these models have real-time RTF and do not add overhead to the RAM or CPU.
In January 2025, the U.S. FTC finalized changes to COPPA, now requiring additional parental consent for data sharing with third-party companies and adding stricter legislative language related to audio data use. 
In view of these enforcements, our study becomes viable towards educational technology especially for the gamification of language development and acquisition, which focuses on assisting speech and language development as well as to encourage engagement and support learning.

\section{Acknowledgements}
This work is supported by NSF Grants 1918032, 2234916, and 2341384; and HPC resources provided by the Texas Advanced Computing Center (TACC).

\bibliographystyle{IEEEtran}
\bibliography{mybib}

\end{document}